\documentclass[aps,prl,reprint,amsmath,amssymb,groupedaddress,showpacs,nobibnotes]{revtex4-1}
\usepackage{graphicx}
\usepackage{bm}     
\usepackage{hyperref} 
\usepackage{dsfont}    
\usepackage{color}

\begin{document}

\title{Scheme for directly observing the non-commutativity of the position and the momentum operators with interference}

\author{Jong-Chan Lee}
\email{spiritljc@gmail.com}
\affiliation{Department of Physics, Pohang University of Science and
Technology (POSTECH), Pohang, 790-784, Korea}

\author{Yong-Su Kim}
\altaffiliation[Present address: ]{Information Technology Laboratory, NIST, Gaithersburg, MD 20899, USA}
\affiliation{Department of Physics, Pohang University of Science and
Technology (POSTECH), Pohang, 790-784, Korea}

\author{Young-Sik Ra}
\affiliation{Department of Physics, Pohang University of Science and
Technology (POSTECH), Pohang, 790-784, Korea}

\author{Hyang-Tag Lim}
\affiliation{Department of Physics, Pohang University of Science and
Technology (POSTECH), Pohang, 790-784, Korea}

\author{Yoon-Ho Kim}
\email{yoonho72@gmail.com}
\affiliation{Department of Physics, Pohang University of Science and
Technology (POSTECH), Pohang, 790-784, Korea}

\date{\today}


\begin{abstract}
Although non-commutativity of a certain set of quantum operators (e.g., creation/annihilation operators and Pauli spin operators) has been shown experimentally in recent years, the commutation relation for the position and the momentum operators has  not been directly demonstrated to date. In this paper, we propose and analyze an experimental scheme for directly observing the non-commutativity of the position and the momentum operators using single-photon quantum interference.  While the scheme is studied for the single-photon state as the input quantum state, the analysis applies equally to matter-wave interference, allowing a direct test of the position-momentum commutation relation with a massive particle.
\end{abstract}



\maketitle

\textit{Introduction.--}
In quantum physics, a certain set of observables do not commute and this non-commutativity of the conjugate observables leads to the uncertainty relation which is at the heart of many unique quantum effects  \cite{Heisenberg27,Bohr28}. It also has been the subject of many illuminating debates on quantum physics \cite{Greenstein06,home}. For instance, the famous Einstein-Bohr debate was on the uncertainty principle regarding the position and momentum measurement as, in quantum physics, two non-commuting observables cannot be measured accurately simultaneously \cite{home}. Moreover, Einstein-Podolsky-Rosen argued against such apparent lack of  \textit{simultaneous physical reality} in their famous 1935 paper \cite{Einstein35}.

Although the commutation relation has been well established theoretically since Heisenberg introduced the canonical commutation relation of the position and the momentum operators, experimental tests on the non-commutativity of conjugate operators have been rather scarce. Non-commutativity of Pauli spin operators ($\sigma_x$, $\sigma_y$, and $\sigma_z$) have been demonstrated with fermions (neutrons) \cite{Wagh97,Hasegawa97} and recently with bosons (photons) \cite{Kim10,Yao10}. Also, recently, non-commutativity of bosonic creation  ($\hat a^\dag$)  and annihilation operators  ($\hat a$) has been demonstrated with photons \cite{Kim08,Zavatta09}. However,  the non-commutativity between the position ($\hat x$) and the momentum ($\hat p$) operators has always been associated with the uncertainty principle regarding the position and the momentum measurement of a particle, as pictured in the Heisenberg microscope \cite{heisenberg,Feynman65}. Note though that the  single-particle uncertainty relation breaks down for the position-momentum entangled two-particle system as discussed in Ref.~\cite{Einstein35} and demonstrated in Ref.~\cite{D'Angelo04,Kim99}. The position-momentum uncertainty relation can be investigated with the single-slit diffraction experiment involving a quantum object \cite{heisenberg,Feynman65} and it has been demonstrated experimentally for neutrons \cite{Shull69} and for large fullerene molecules \cite{Nairz02}. It is nevertheless interesting to point out that, to the best of our knowledge,  the non-commutativity relation for the position and the momentum operators itself has not been directly observed to date as was demonstrated for  the Pauli spin operators and the bosonic creation/annihilation operators. 

In the present work, we propose and analyze an experimental scheme to directly observe non-commutativity of the position and the momentum operators using the transverse spatial degree of freedom $x$ of a single-photon wave function $\psi(x)$ (in the sense that $|\psi(x)|^2$ gives the probability distribution) \cite{note}.  The position and the momentum operators $\hat x$ and $\hat p$ are implemented using position-dependent attenuator, phase plates, and lenses. The commutator and the anti-commutator of the position and the momentum operators are constructed with single-photon quantum interference. For an initial Gaussian wave function $\psi(x)$, we find that applying the commutator leaves the state unchanged whereas applying the anti-commutator results a Wigner function with negativity, starkly different from the Wigner function of the initial wave function \cite{Mukamel03}. Finally, we discuss how the proposed scheme can be applied to matter-wave interferometry  to directly observe the non-commutativity of the position and the momentum operators for a particle with mass or a macroscopic quantum state of matter.


\textit{Implementing $\hat x$-$\hat p$ commutation operations.--} 
Consider a  quasi-monochromatic single-photon traveling in the $z$-direction.  Since the Hilbert space that represents the transverse spatial degrees of freedom  of the photon is isomorphic to the Hilbert space that describes the quantum state of a point particle in two-dimension \cite{Stoler81,Tasca11}, we may use the quantum wave function formalism for a point particle to describe the transverse spatial wave function of a single-photon. We choose to describe only one  transverse spatial degree of freedom, namely the transverse position of the photon $x$, without loss of generality due to the orthogonality.

In the position basis, the relation $\hat x |x\rangle = x|x\rangle$ holds so that an arbitrary pure state $|\alpha\rangle$ can be written as
$|\alpha\rangle=\int dx\,\psi_{\alpha}(x)|x\rangle$, where $\psi_{\alpha}(x)$ is the transverse  spatial wave function for the  state $|\alpha\rangle$. In the conjugate momentum basis, the relation $\hat p |p\rangle = p|p\rangle$ holds  so that $|\alpha\rangle=\int dp\,\phi_{\alpha}(p)|p\rangle$, where $\phi_{\alpha}(p)$ is the corresponding wave function for $|\alpha\rangle$ in this basis.

\begin{figure}[t]
\centering
\includegraphics[width=3.2 in]{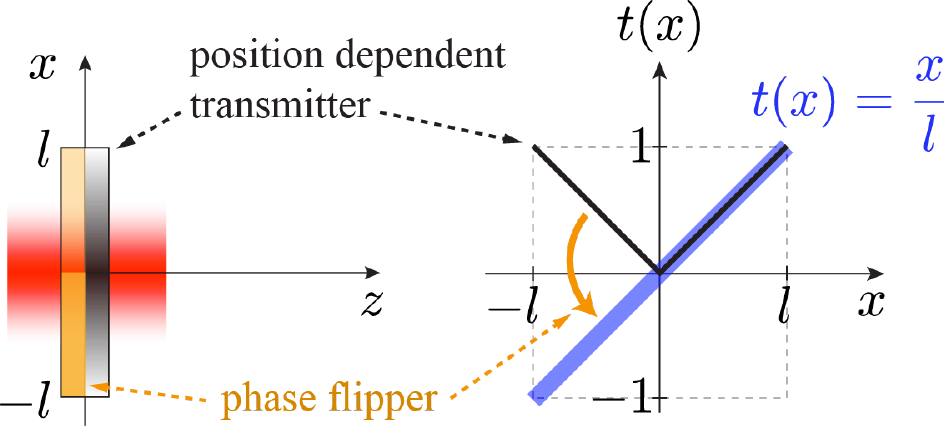}
\caption{Scheme for implementing $\hat x$ operation. Applying $\hat x$ to the wave function  introduces the amplitude transmission coefficient $t(x) = x/l$. The phase shifter introduces the relative phase shift of $\pi$ for the region $x \in [-l,0]$. The position dependent transmitter introduces the linear amplitude transmission coefficient $t(x)=|x|/l$. } \label{position}
\end{figure}

To implement a quantum operator is to find a quantum operation which results the desired output quantum state. For the position operator $\hat x$,
\begin{equation}
\hat{x}|\alpha\rangle=\hat{x}\left(\int dx\,\psi_{\alpha}(x)|x\rangle\right)=\int dx\,x\psi_{\alpha}(x)|x\rangle,
\end{equation}
which means that the action of  $\hat x$ to the wave function $\psi_\alpha(x)$ is multiplication of $x$ to the wave function, i.e.,  $x \psi_\alpha(x)$. Then, the corresponding operation is to introduce the amplitude transmission coefficient $t=x/l$ and it can be implemented by using a set of a $\pi$ phase shifter followed by an attenuator with the linear amplitude transmission coefficient $t(x)=|x|/l$ , see Fig.~\ref{position}. The phase shifter introduces the relative phase shift of $\pi$ for the region $x \in [-l,0]$ with respect to the region $x \in [0,l]$. The phase shifter can be implemented, for instance,  with a piece of glass by polishing away thickness corresponding to $\lambda/2$ for the region  $x \in [-l,0]$, where $\lambda$ is the  central wavelength of the photon. The phase flipper and the attenuator together  achieve the overall amplitude transmission coefficient of $t(x)=x/l$, implementing the operation corresponding to a dimensionless position operator $\tilde{x}=\hat{x}/l$.

 

\begin{figure}[t]
\centering
\includegraphics[width=3.2 in]{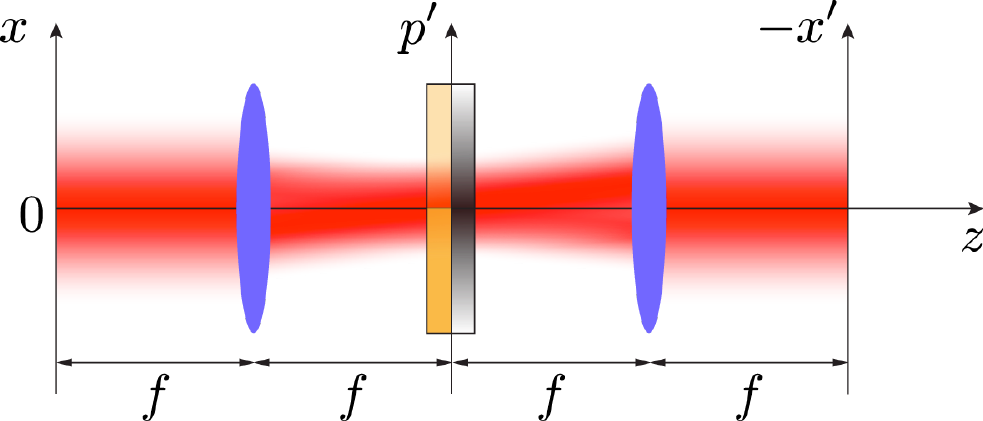}
\caption{Scheme for implementing $\hat p$ operation. Two lenses of the same focus $f$ make a $4f$ imaging system with the $\hat x$ operation implemented at the Fourier plane.  
} \label{momentum}
\end{figure}

For the momentum operator $\hat p$, we have a similar result,
\begin{equation}
\hat{p}|\alpha\rangle=\hat{p}\left(\int dp\,\phi_{\alpha}(p)|p\rangle\right)=\int dp\,p\phi_{\alpha}(p)|p\rangle.
\end{equation}
To implement the quantum operation which results $p\phi_{\alpha}(p)$, we simply need to place the set consisting of the phase shifter and the attenuator in Fig.~\ref{position} at the Fourier plane of the incoming photon.  This is easy to see by simply re-expressing the above equation in the position-basis, 
$
\hat{p}|\alpha\rangle =\int dp\,p\phi_{\alpha}(p)|p\rangle = \int dx  \langle x|\hat{p}|\alpha\rangle|x\rangle
$. 
The scheme for implementing $\hat p$ operation is shown in Fig.~\ref{momentum}. Two lenses of the same focus $f$ form a $4f$ imaging system and the $\hat x$ operation is implemented at the Fourier plane of the $4f$ imaging system.




To be more specific, consider the scheme in Fig.~\ref{momentum} and assume that the wave function in the input plane ($z=0$) is given as $\psi_{\alpha}(x)$. The first lens maps the position-space wave function $\psi_{\alpha}(x)$ into the momentum-space wave function  $\phi'_{\alpha}(p')$ \cite{Goodman05},
\begin{equation}
\phi'_{\alpha}(p')=\frac{1}{\sqrt{i\lambda f}}\int\psi_{\alpha}(x)\exp\Big(-i\frac{2\pi}{\lambda f}x p'\Big)dx,
\end{equation}
where $p'=\frac{\lambda f}{2\pi \hbar} p$. At the Fourier plane, the set of the phase shifter and the attenuator performs the transformation $\phi'_{\alpha}(p') \rightarrow  p' \phi'_{\alpha}(p')/l$. Finally, at the output of the optical system, i.e., $z=4f$, the wave function is now given as
\begin{equation}
\frac{1}{\sqrt{i\lambda f}}\int\frac{p'}{l\,}\phi'_{\alpha}(p')\exp\Big(i\frac{2\pi}{\lambda f}x' p'\Big)dp'=-i\langle x'|\tilde{p}|\alpha\rangle,
\end{equation}
where $\tilde{p}=\frac{\lambda f}{2\pi \hbar l}\hat{p}$ is the dimensionless momentum operator. Note that we chose $x'=-x$ to account for the inverting nature of a $4f$ imaging system.

\begin{figure}[t]
\centering
\includegraphics[width=3.2 in]{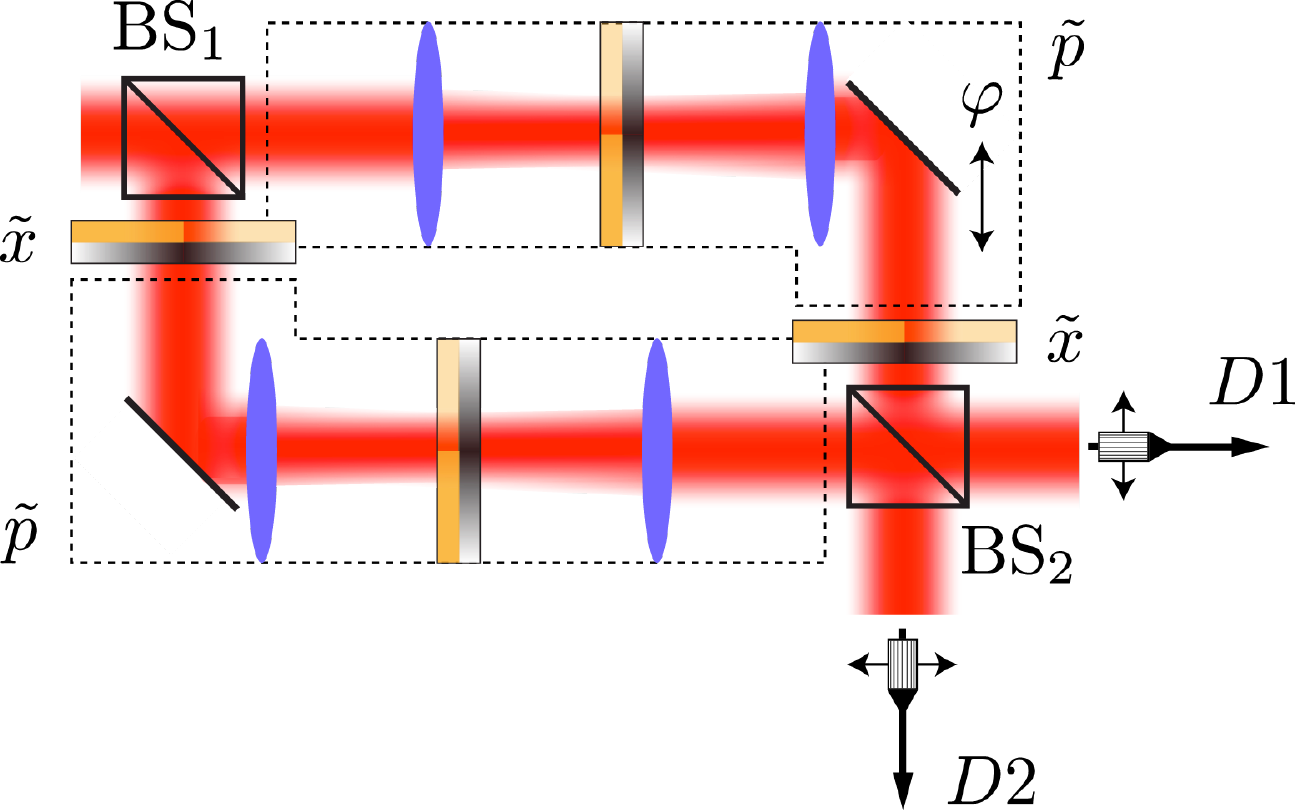}
\caption{Coherent superposition of two quantum operations $\tilde{x}\tilde{p}$ and $\tilde{p}\tilde{x}$ is accomplished with a Mach-Zehnder interferometer. Single-mode fiber tips (connected to single-photon detectors) are scanned in transverse directions to measured the output states.
 } \label{interferometer}
\end{figure}

We have so far discussed how to implement the dimensionless position $\tilde x$ and momentum $\tilde p$ operators. Let us now discuss how to probe the commutation relations for the operators by using single-photon interference. Consider the experimental scheme shown in Fig.~\ref{interferometer}. A single-photon state enters the Mach-Zehnder interferometer  via the beam splitter at left. At each arm of the interferometer, optical systems which implement the dimensionless position and momentum operators are placed but in different order. In the upper path, $\tilde p \tilde x$ is implemented and in the lower path $\tilde x \tilde p$ is implemented. The two quantum operations are then coherently superposed at the second beam splitter with a relative phase $\varphi$ \cite{Kim08,Zavatta09}. Quantum superposition of operators are implemented at the output of the Mach-Zehnder interferometer, namely, $\tilde{x}\tilde{p}+ e^{i\varphi}\tilde{p}\tilde{x}$ at $D_1$  and $\tilde{x}\tilde{p}- e^{i\varphi}\tilde{p}\tilde{x}$ at $D_2$. The resulting wave functions can be analyzed by measuring the single-photon detection probabilities with scanning fiber tips in the transverse direction.

\textit{Results and Discussions.--}
Suppose now that a single-photon with the wave function $\psi(x)$ enters the interferometer   and the relative phase is set at $\varphi = \pi$.  At $D_1$, we have $[\tilde x, \tilde p] \psi(x) = \frac{C}{\hbar}[\hat{x},\hat{p}]\psi(x) $, where $C \equiv \frac{\lambda f}{2\pi l^2}$. Since $[\hat{x},\hat{p}]=i\hbar$,  the commutator for the dimensionless operators becomes $[\tilde x, \tilde p] =i C$ so that the commutator operated on the wave function leaves the state unchanged (after normalization). Similarly, at $D_2$, the anti-commutator for the dimensionless operators acted on the wave function results  $\{\tilde x, \tilde p\} \psi(x) = \frac{C}{\hbar}\{ \hat x, \hat p\} \psi(x)$ which means that the resulting wave function appearing at $D_2$ is different from the input wave function. Note that the commutator and the anti-commutator output ports, set at $D_1$ and $D_2$ respectively, can be switched by choosing the relative phase value  $\varphi = 0$. 

\begin{figure}[t]
\centering
\includegraphics[width=3 in]{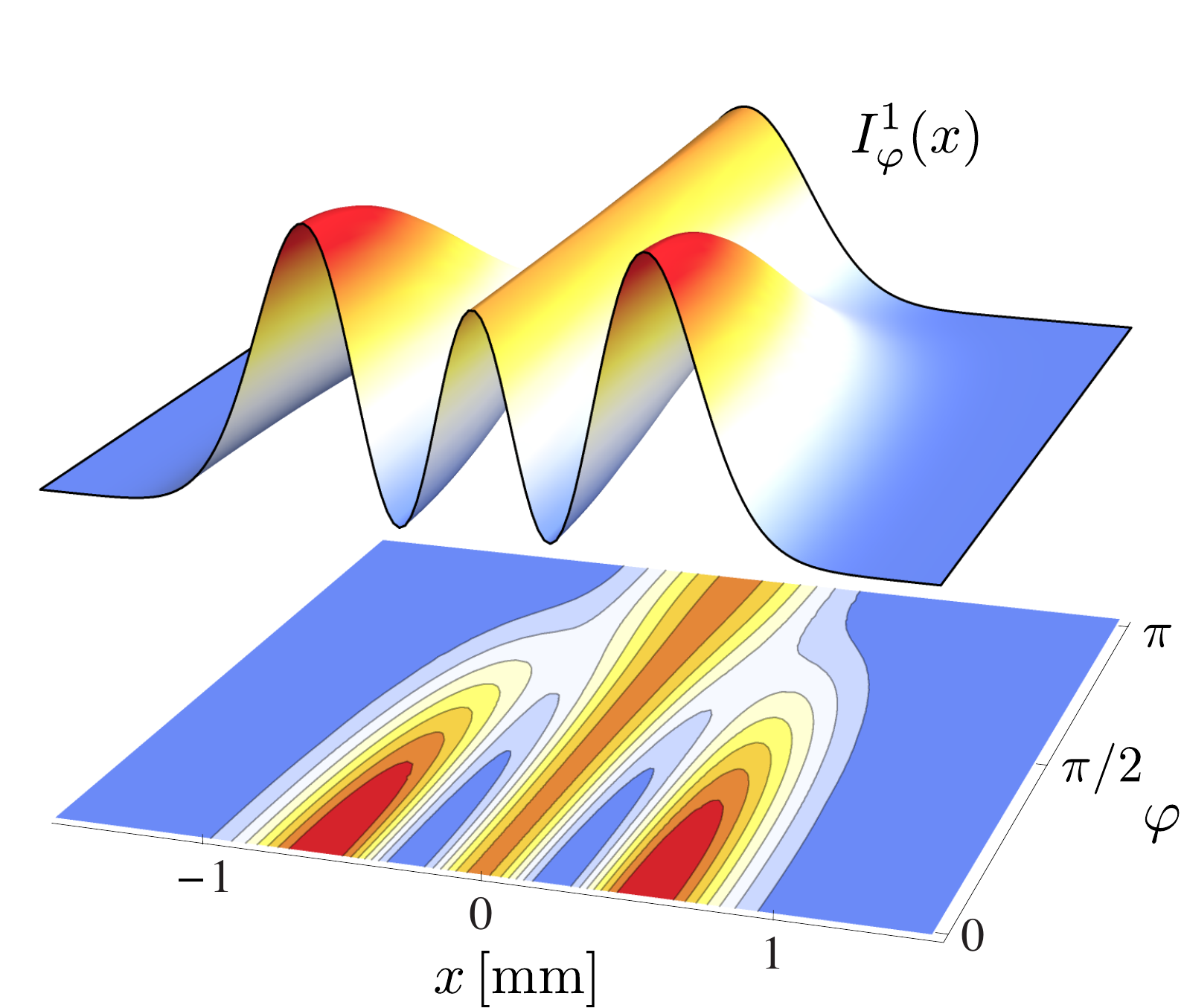}
\caption{Probability distribution $I^1_{\varphi}(x)$ at $D_1$ calculated as a function of $\varphi$ and $x$ assuming a Gaussian input wave function. The commutator $\varphi=\pi$ acted on the input Gaussian wave function leaves the wave function unchanged. The anti-commutator $\varphi=0$, however, causes the input wave function to change, resulting interference. 
} \label{superposition}
\end{figure}

As an example, let us consider a quasi-monochromatic single-photon state with a Gaussian wave function 
\begin{equation}
\psi(x)=\sqrt[4]{\frac{2}{\pi w^2}}\exp\left(-{x^2}/{w^2}\right)
\end{equation}
at the input port of the Mach-Zehnder interferometer. The detection probability $I_{\varphi}^1(x)$ at $D_1$ as a function of  $x$ and $\varphi$ is shown in Fig.~\ref{superposition}. Here we have assumed $\lambda= 800$ nm, $w = 0.5$ mm, $l = 1.5$ mm, and $f=50$ cm. The figure shows that, at $\varphi=\pi$ which corresponds to the commutator case, the Gaussian input wave function is reproduced at the output as expected. On the other hand, at $\varphi=0$ which corresponds to the anti-commutator case, the output shows interference fringes along the $x$ direction.


\begin{figure}[t]
\centering
\includegraphics[width=3.2 in]{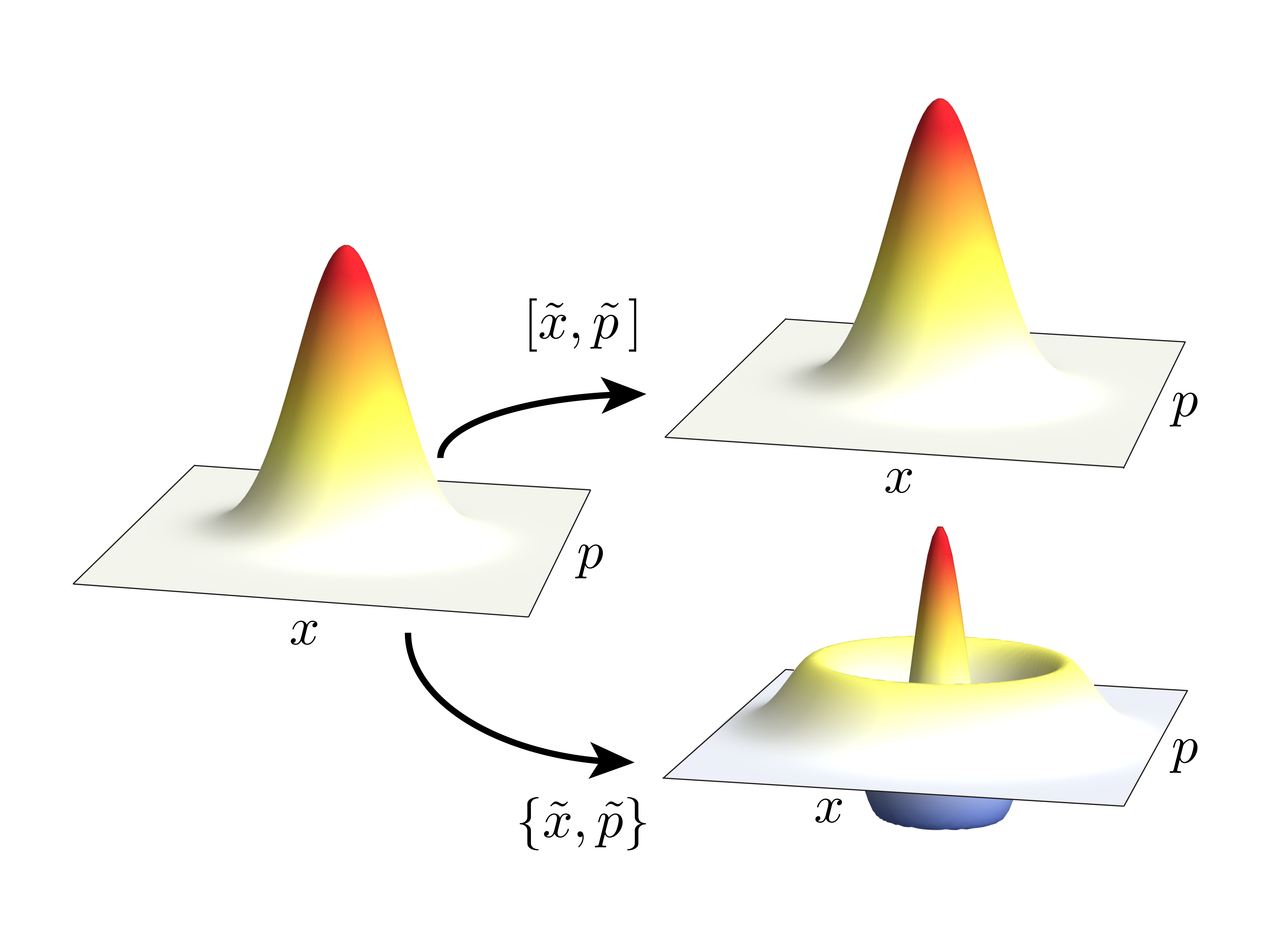}
\caption{The spatial Wigner functions $W(x, p)$ of the input wave function $\psi(x)$ (Gaussian; everywhere positive) and the states after operating commutator $[\tilde x, \tilde p]$ and anti-commutator $\{ \tilde x, \tilde p\}$. Applying the commutator to the wave function results in the identical spatial Wigner function as the input, while applying the anti-commutator lead to a starkly different Wigner function with negativity.} \label{wigner}
\end{figure}

A complete characterization of the spatial coherence of the wave function $\psi(x)$, however, requires tomographic reconstruction of the spatial Wigner function $W(x,p)$ where  $x$ and $p$ refer to actual position and momentum of the single-photon.  The spatial Wigner function $W(x,p)$  can be reconstructed by employing, e.g., an area-integrated detection scheme \cite{Mukamel03}. In Fig.~\ref{wigner}, the spatial Wigner functions for the input wave function and the two output wave functions (i.e., the commutator-operated wave function  and the anti-commutator-operated wave functions) are shown. The figures clearly show that the input and the commutator-operated wave functions have the same Wigner functions. In other words, it shows that the quantum operation corresponding to the $\hat x$ and $\hat p$ commutator is indeed equivalent to (a constant multiple of) the identity operation. On the other hand, the anti-commutator-operated wave function  exhibits a completely different Wigner function, interestingly with a clear signature of negativity. 



 

\textit{Conclusion and Outlook.--}
The commutation relation for the position and the momentum operators leads to the position-momentum uncertainty relation and, in experiment, the position-momentum uncertainty relation has been demonstrated in single-slit diffraction experiments involving a massive (neutrons, fullerene, etc) or a massless particle (photons) \cite{Shull69,Nairz02,Feynman65}. In this work, we have proposed and analyzed an interferometric scheme for directly (i.e., without involving the uncertainty relation) observing the commutation relation for the position and the momentum operators using single-photon quantum interference. The proposed scheme requires only linear optical elements and single-photon detectors so it should be possible to realize such an experiment if the position-dependent attenuator and the $\pi$ phase shifter  can be precisely engineered. In practice, a soft-edge graduated neutral density filter (whose amplitude transmission coefficient is linear) can function as a position-dependent attenuator and a molding technique can be used for producing a precise phase shifter \cite{oem}.


Although the interferometric scheme proposed in this paper is focused on single-photon interferometry, the proposed concept can readily be expanded to matter-wave interferometry involving a massive particle or even the macroscopic quantum state of matter. For instance, essential elements in the proposed scheme can be built for a Bose-Einstein condensate, a macroscopic quantum object. Position-dependent attenuators can be built using the quantum tunneling effect through a laser-induced potential barrier \cite{duan10}, focusing of a matter wave can be accomplished by using light as an atomic lens \cite{Timp92,Dubetsky98}, and a matter-wave interferometer can be built by using the sequential Bragg momentum transfer effect \cite{Chiow11}. Moreover, it is possible to reconstruct the spatial Wigner function for a massive particle \cite{Kurtsiefer97}, thus making a direct experimental test of the position-momentum commutation relation with a macroscopic quantum object within the reach of the present-day technology.



This work was supported in part by the National Research Foundation of Korea (2011-0021452 and 2012-002588). J.C.L and H.T.L. acknowledge support from the National Junior Research Fellowships (2012-000741 and 2012-000642, respectively).



\begin{thebibliography}{}
\bibitem{note} For a photon, $\psi(x)$  does not satisfy the Schroedinger equation as it should represent a real electric field satisfying the Maxwell's equations.
\end{thebibliography}
\end{document}